\documentclass[preprint]{vgtc}               % preprint
\graphicspath{{figures/}{pictures/}{images/}{./}} % where to search for the images

\usepackage{times}                     % we use Times as the main font
\usepackage{tabu}                      % only used for the table example
\usepackage{booktabs}                  % only used for the table example
\usepackage{lipsum}                    % used to generate placeholder text
\usepackage{mwe}                       % used to generate placeholder figures

\usepackage{mathptmx}                  % use matching math font

\usepackage{soul,color}
\soulregister\cite7
\soulregister\ref7
\soulregister\pageref7
\renewcommand{\hl}[1]{#1} %%for undo highlighting

\onlineid{1125}

\vgtccategory{Empirical Study}

\vgtcinsertpkg

\title{Exploring AR Label Placements in Visually Cluttered Scenarios}

\author{Ji Hwan Park\thanks{e-mail: jhpigm@rit.edu}\\ %
    \scriptsize Rochester Institute of Technology %
\and Braden Roper\thanks{e-mail:bradenroper@ou.edu}\\ %
    \scriptsize University of Oklahoma %
\and Amirhossein Arezoumand\thanks{e-mail:amirhossein.arezoumand@ou.edu}\\ %
    \scriptsize University of Oklahoma %
\and Tien Tran\thanks{e-mail:tien.g.tran@ou.edu}\\ %
\scriptsize University of Oklahoma
}

\teaser{
  \centering
  \includegraphics[width=\linewidth]{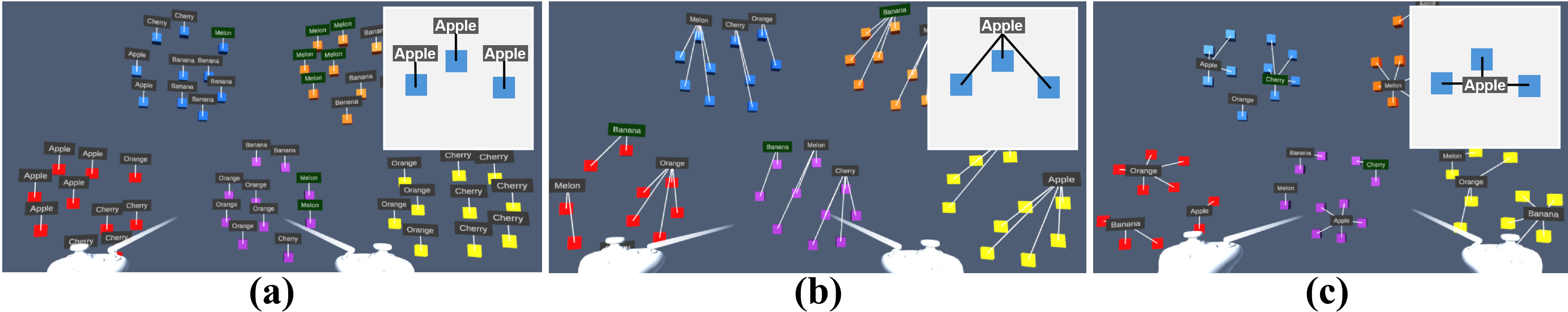}
   \vspace{-8 mm}
  \caption{Examples of three label placements: (a) Situated Individual, (b) Situated Grouped, and (c) Centered Grouped in simulated AR views \hl{and the diagrams of them (top right). Each cube's color shows its spatial region, not an item type for the Identify task.}}
  \label{fig:teaser}
}

\abstract{
          We investigate methods for placing labels in AR environments that have visually cluttered scenes. As the number of items increases in a scene within the user' FOV, it is challenging to effectively place labels based on existing label placement guidelines. To address this issue, we implemented three label placement techniques for in-view objects for AR applications. We specifically target a scenario, where various items of different types are scattered within the user's field of view, and multiple items of the same type are situated close together. We evaluate three placement techniques for three target tasks. Our study shows that using a label to spatially group the same types of items is beneficial for identifying, comparing, and summarizing data. 
} % end of abstract

\keywords{Label placement, view management systems, augmented reality, situated visualization.}

\begin{document}



\firstsection{Introduction}
\maketitle
Augmented reality (AR) is a technology that combines computer-generated graphics with the physical world. This technology overlays digital information and annotations onto real-world scenes and objects, providing users with extra information about their surroundings. AR can be used to improve users' understanding of real-world situations and help them complete tasks more efficiently.
For instance, an AR tool can assist users in identifying books in a specific category within a library or comparing books based on their ratings at a bookstore.

In AR, labeling is a popular method for providing information related to physical objects/items, linking the information to the items by using leader lines. Label layout design can affect users' performance in several tasks in AR~\cite{Azuma:2003, Madsen:2016}. Many label placement methods have been proposed to find optimal label locations in different AR scenarios, such as labeling out-of-view objects~\cite{Lin:2023}, dynamic scenarios~\cite{Chen:2023}, and finding a specific label~\cite{Zhou:2021}. The general principles of these label placement methods are 1) avoiding leader line crossing, 2) minimizing the length of leader lines, and 3) avoiding/reducing the overlap of labels with associated objects and other labels. 
Most existing research has focused on spatially sparse scenarios, where the number of objects is usually less than 30 in users' field-of-view (FOV). However, in real-life scenarios, there can be a large number of objects in the AR FOV, which can cause visual clutter in the FOV~\cite{Azuma:2001}.

To address this issue, we investigate how to place labels in visually cluttered scenes in AR.
We specifically focus on a scenario where several groups of items/objects are densely distributed in users' FOV, and multiple items of the same type are located in close proximity to one another. For example, several copies are placed together at different locations in a library. 
We aim to study the impact of using a representative label on items of the same type and the location of a representative label on reducing visual clutter and helping users perform tasks.
For this, we designed three conditions: \textit{Situated Individual, Situated Grouped, and Centered Grouped}. In the \textit{Individual}, we place one label for its associated item, while in the \textit{Grouped}, we use one label for a group of the same items that are spatially close together. For the \textit{Situated}, we position a label above the top of an item/group, and for \textit{Centered}, we locate a label at the center of spatially grouped items. 
We evaluated three label placement methods through a user study. 
In the study, we implemented and used a virtual reality (VR) application to simulate an AR environment as it allows us to easily configure visually cluttered environments~\cite{Lin:2023}.
We found that a label placement method that is \textit{Situated} and \textit{Grouped} helps users perform tasks in the study faster. 
Additionally, many participants preferred the \textit{Grouped} placement methods over the \textit{Individual} method. 

\section{Related Work}

\subsection{Situated Visualization/Analytics in AR}
Situated Visualization/Analytics (SVA) in AR refers to projecting information onto physical space through AR. This approach aims to display information and visualizations in a way that is sensitive to the surrounding environment, relevant to the context, and enhances the user's understanding and interaction with both the data and their physical surroundings~\cite{bressa2021situated,elsayed2015situated}. 
% Bressa et al.~\cite{bressa2021situated} extended this concept by looking at five perspectives: space, time, place, activity, and community. 

SVA offers several advantages, such as the proximity of visual information/view to their corresponding objects and ease of locating a view for each potential object. However, it becomes challenging to compare different views. To solve this issue, Wen et al.~\cite{Wen2022EffectsOV} proposed a cylindrical framework for multiple views, using a force-directed method to optimize distances between views, physical objects, and users. Huynh et al.~\cite{8798358} proposed an in-situ labeling for language learning. To enhance word learning, they proposed a client-server architecture that allows for real-time object labeling on the objects in an AR scenario. An object detection algorithm is employed to initially detect objects in the scene, and subsequently, labels are placed on these objects based on user eye-gaze information. 
Whitlock et al.~\cite{Whitlock:2020} investigated the effect of display types and visual encoding channels in interpreting data visualization in SVA.
Lee et al.~\cite{Lee:2024} explored the design space and proposed some guidelines for SVA. Among them, they suggested that the high density of referred items should be considered in SVA design.     
Because labeling is one of the important tasks in SVA~\cite{Lee:2024},
our findings will be integrated into SVA and aid users in exploring and understanding data in AR with the high density of objects/items.

% They adopt the Grasset et.al~\cite{grasset2012image} method and create a penalty function using saliency and edge detection information from the frames. Subsequently, labels move to candidate areas with smaller penalty functions. By applying this method, they ensure that the labels are placed in less important areas, which leads to important objects not being missed by the users.

\subsection{Grouping Labels}
Labels for complex objects or scenes often cause visual clutter or occlude structures~\cite{Tatzgern:2013}. To address these issues, grouping methods have been proposed.
Tatzgern et al.~\cite{Tatzgern:2013} proposed a label placement method to reduce visual clutter. They clustered labels based on the text similarity and displayed a representative label for each cluster. 
Fink et al.~\cite{Fink:2012} also used a clustering method for labeling sites to reduce visual clutter. They grouped sites and showed a site for each cluster that represents the spatial distribution of all sites.
M{\"{u}}hler and Preim~\cite{Mühler:2009} used a single label if target items have the same structure and they are close. However, they didn't discuss where to place a single label for grouped items, nor did they evaluate the effectiveness of their grouping method.
G{\"{o}}tzelmann et al.~\cite{Gotzelmann:2006b} proposed a method to show the groups of labels, grouping and placing labels based on user-defined ontology.
All these methods are potentially useful to reduce visual clutter, but they haven't been evaluated in AR/VR environments. We evaluate three label placement methods, including two grouping methods, in a simulated AR environment to understand the effects of grouping methods.

\subsection{Label Placement in AR}
Labels have been widely used in visualization to display information associated with objects in AR.
Azuma and Furmanski~\cite{Azuma:2003} explored 2D label placement algorithms in AR where objects are close to each other.
Visual saliency-based label placement methods were proposed to show important objects with minimum occlusion~\cite{Grasset:2012,Hegde:2020,Rakholia:2018,Jia:2021}. 
Grasset et al.~\cite{Grasset:2012} also adaptively changed leader lines, anchor, background, and text for optimal label placement.
Bell et al.~\cite{Bell:2001} presented a system to display 3D models and their associated labels. They placed labels to avoid overlapping 3d models or other labels.
Tatzgern et al.~\cite{Tatzgern:2014} introduced a 3D geometry of labels-based layout to show the smooth label motion during camera movement.
Tatzgern et al.~\cite{Tatzgern2016AdaptiveID} aggregated information by using hierarchical clustering and displayed aggregated data to avoid visual clutter based on user preferences.
Madsen et al.~\cite{Madsen:2016} compared four methods, including the 3D geometry of labels-based layout, and found that the 3D geometry of labels-based layout with a limited update rate is good for locating annotations.

Gebhardt et al.~\cite{Gebhardt:2019} used a reinforcement-learning method to filter labels based on users' gaze patterns.
Köppel et al.~\cite{Koppel:2021} introduced a context-responsive method for AR label placement by reducing occlusion, controlling the level of detail of labels, and using a smooth transition.
Zhou et al. ~\cite{Zhou:2021} proposed a method for partially sorting label placement, where they found the optimal number of circles and positioned labels on those circles to minimize long and crossing leader lines. 
RL-label~\cite{Chen:2023} is a reinforcement learning-based AR label placement method for dynamic scenes. 
This method reduces overlap and leader line crossing, and is easy to follow labels, as compared to existing methods.
Lastly, Lin et al.~\cite{Lin:2023} explored the design space of AR labels and evaluated five different label placement approaches for out-of-view objects for situated analysis.
While these methods focused on spatially sparse scenarios, we focus on a visually cluttered environment with different types of items, where groups of identical items are scattered throughout the user's FOV.

\section{Design Space and Tasks}
% A label is a visual representation that provides information associated with a target object/item. In this section, we discuss our design considerations, target tasks, and a usage scenario in a visually cluttered AR environment.

\subsection{Design Space}
When we place labels, we need to consider several aspects to have better legibility and aesthetics, including reducing crossing leader lines and minimizing overlaps between objects and labels~\cite{Bekos:2019}. While all these aspects are important when designing a label placement method, we consider two design aspects: the level of visual clutter and the lengths of leader lines because those can affect users' performance in our target environment~\cite{Grasset:2012}.
% we focus on the following two properties to understand the effects of the properties in AR with many objects.

\vspace{5pt}
\noindent
\textbf{Visual Clutter.} \,
If there is a large amount of information associated with items in an environment, it makes the environment visually cluttered. By reducing the number of labels or grouping the same object, we can minimize visual clutter~\cite{Tatzgern:2013,Mühler:2009}. Reducing visual clutter may help users perform a visual search task~\cite{Verghese:2004}. 

\vspace{2pt}
\noindent
\textbf{Lengths of Leader Lines.} \,
One of the design principles for label placement is that labels should be as close as possible to their object~\cite{Vollick:2017}. However, short leader lines sometimes make it challenging to find labels and compare them~\cite{Lin:2023}. 

\subsection{Tasks}
Based on previous work~\cite{Lin:2023,Chen:2023}, we identified three target tasks to search for items of interest in AR.

\vspace{5pt}
\noindent
\textbf{T1. Identify: Identify a single target group. } \,
Users want to identify a group of items that are spread all over the place. For this, users need to memorize the type and location of each item. 
Labels in AR can provide spatial cues of the target items to help users identify them~\cite{Lin:2023}.

\vspace{2pt}
\noindent
\textbf{T2. Compare: Compare multiple items in a target group. } \,
Users need to compare data values associated with multiple items in a single target group, which are placed in different locations.  This task is more difficult than the Identify task because users need to memorize the data values of the items as well.
Labels in AR can allow users to compare multiple items visually, reducing the burden of memorizing the items’ data values~\cite{Lin:2023}. 

\vspace{2pt}
\noindent
\textbf{T3. Summarize: Summarize all items in all target groups. } \,
Lastly, users need to extend their target items from one group to multiple groups. This task is the most challenging because users need to summarize all items in each target group.  
Using labels, users can summarize the overall distribution of items or determine outliers across all target groups~\cite{Brehmer:2013}.

\section{Design}
Based on our design considerations, we design three label placement methods (Fig.~\ref{fig:teaser}) for in-view items, where groups of the same items are distributed within the user's FOV. 
In the Situated Individual, we place a label for each referred item. \hl{The Situated Grouped and Centered Grouped create item groups based on their type and spatial proximity, and use a representative label for each group.}   

\vspace{4pt}
\noindent
\textbf{Situated Individual.} \,
Situated Individual is the most popular label placement approach in AR/VR applications~\cite{Lin:2023}.
This method results in less clutter on leader lines compared to placing labels at the bottom of an AR view~\cite{Lin:2023}.
Thus, we used situated individual labels as a baseline method for our study (Fig.~\ref{fig:teaser} (a)).   
We follow Lin et al.'s Situated label design~\cite{Lin:2023}.
We place a label directly above its target item with a straight leader line linking the label's bottom center to the item's center. To avoid overlaps, we adjust label positions vertically if they are too close to items or each other.

If a few items are present, this method may produce short leader lines due to items being distantly placed, and/or labels can be near items without overlaps. Once users find a label of interest, they may easily locate the associated item because it is below the label.
However, in our target environments, many items are close together, so leader lines may be long.
Moreover, although this method is familiar to users, it may be challenging due to numerous labels causing clutter, especially with many similar items nearby.

% \begin{figure}
%      \centering
%      \begin{subfigure}{0.49\linewidth}
%          \centering
%          \includegraphics[width=\textwidth]{figures/SI_ex.png}
%          \caption{}
%          \label{fig:y equals x}
%      \end{subfigure}
%      \hfill
%      \begin{subfigure}{0.49\linewidth}
%          \centering
%          \includegraphics[width=\textwidth]{figures/SG_ex.png}
%          \caption{}
%          \label{fig:three sin x}
%      \end{subfigure}
%         \caption{Examples of a scene with long leader lines in (a) Situated Individual and (b) Situated Grouped.}
%         \label{fig:long_leaderlines}
% \end{figure}
   
\vspace{4pt}
\noindent
\textbf{Situated Grouped.} \,
In some situations, you may find multiple similar items close together - for instance, fruits in a grocery store or books in a library. In these cases, it may be necessary to display a representative label for each spatial group of items of the same type to reduce visual clutter and label overlap.

To achieve this, we use the Situated Grouped method (Fig.~\ref{fig:teaser} (b)), where one label represents multiple items forming a spatial cluster/group.
\hl{For each type, we group items with the same type and ratings, if available, by using DBSCAN~\cite{Ester:1996} with a user-defined Euclidean distance. Users can adjust this distance based on the level of visual clutter or spatial distribution.}
For each group, we place a label above it, linking leader lines to items to minimize overlaps, similar to the Situated Individual method. This method can be familiar to users due to its similarity to the Situated Individual method and reduces label overlap. Similar to the Situated Individual method, users may also easily identify items linked to a label as they are listed below it. 
However, if multiple vertically spread items of the same type are grouped, the leader line may be long because a single label is placed above the entire group.

\vspace{4pt}
\noindent
\textbf{Centered Grouped.}
Another method to minimize visual clutter is the Centered Grouped (Fig.~\ref{fig:teaser} (c)). In this method, a single label is placed at the center of an item group that is detected using the same approach as that used on the Situated Grouped. 
Sometimes, this central label can be too close to items, leading to overlaps. To address this, we use a force-directed approach~\cite{Bekos:2019}, resulting in shorter leader lines that minimize gaze movement, compared to the situated grouped placement.
However, identifying objects linked to a label can be more difficult than with the Situated Grouped method, as items may be spread around the label, complicating users' predictions of item locations.

Based on our literature survey and our analysis of the label placement methods, we had two main hypotheses:

\vspace{5pt}
\noindent
\textbf{H1.} The Situated Grouped and Centered Grouped placements will perform better than the Situated Individual placement in terms of completion time.

% \vspace{1pt}
\noindent
\textbf{H2.} The Centered Grouped placement will perform better than the Situated Grouped placement in terms of completion time.

\section{User Study}
% We conducted a controlled study to understand the usefulness of three label placement methods in visually cluttered scenes.  

\subsection{Experimental Design}

\vspace{5pt}
\noindent
\textbf{Dataset.} \,
We generated a dataset with 50 objects/items. We divided a user's FOV into 5 regions and randomly distributed 10 items per region. The items in each region were further divided into subgroups. Each subgroup could have at most 5 items and was assigned to a different class/type. Subgroups (item groups) in each region couldn't have the same type. Each type had a unique fruit name (e.g., apple, orange) and was randomly assigned a rating between 1 and 5. \hl{The ratings were displayed directly above their labels (e.g.,} \raisebox{-.2ex}{\includegraphics[height=2ex]{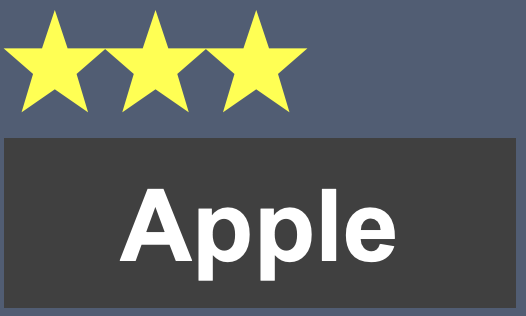}}\hl{)}. There were 5 types in total.

% \begin{figure}[t]
%   \centering
%   \includegraphics[width=0.8\linewidth]{figures/layout.png}
%   \caption{%
%     An example of object/item distribution
%   }
%   \label{fig:layoutdesign}
% \end{figure}

% \vspace{5pt}
% \noindent
% \textbf{Condition.} \,
% We evaluated three label placement conditions described in Sec.4: \textit{Situated individual, Situated Grouped, and Centered Grouped}.  
% All labels were displayed on a semi-transparent AR canvas to mimic an actual AR headset.  
% These conditions were implemented by using Unity~\cite{unity}, where all labels were displayed on a semi-transparent AR canvas to mimic an actual AR headset (Microsoft HoloLens). 

\vspace{5pt}
\noindent
\textbf{Experiment Set-Up.} \,
To easily manipulate the distribution of items, we use a VR headset to simulate the real-world object space and AR screen, which has been used in several AR studies~\cite{Chen:2023,Lin:2023,Marquardt:2020}.
We used an HTC VIVE Pro virtual reality headset, which has a 110\textdegree FoV and 1440 × 1600 pixels per eye to display items and their labels on a semi-transparent simulated AR screen in VR. The headset was connected to a desktop PC with Intel Xeon Silver 4210R 2.40GHz processor and NVIDIA A6000 graphics card.
The study required no head movement because all the items were displayed within a participant's FOV.

\subsection{Task}

We used three tasks described in Section 3.2 that are common in visualization studies~\cite{Brehmer:2013} and AR label studies~\cite{Lin:2023,Chen:2023} to understand the effects of three label placement methods.

\begin{figure}[t]
  \centering
  \includegraphics[width=\linewidth]{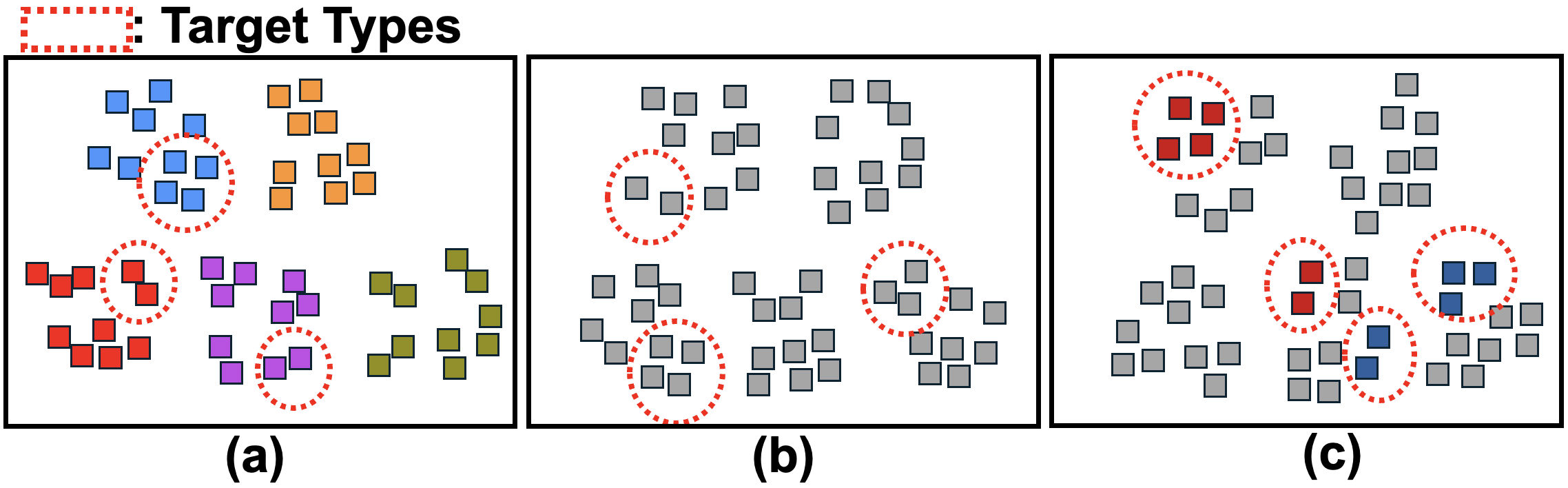}
   \vspace{-8 mm}
  \caption{%
    Examples of target types in our three tasks: (a) Identify, (b) Compare, and (c) Summarize. In (a) and (b), we have only one target type (circles with a dotted line). In (c), we have two target types (colored blue and red in circles with a dotted line).
  }
   \vspace{-4 mm}
  \label{fig:tasks}
\end{figure}

\vspace{5pt}
\noindent
\textbf{Identify.} \,
\textit{How many item colors does a type associated with green labels have?}
There was one target type with green labels, while the other labels were gray. Each region was assigned a unique color, resulting in 5 colors for items. Participants identified how many spatial subgroups/clusters of the target type existed, varying from 2 to 5 (Fig.~\ref{fig:tasks} (a)).
% There was one target type. The color of the labels for the target type was green, and the color of the other labels was grey. Additionally, each region has been assigned a unique color, and items in each region were colored in the same assigned color. Thus, five colors are used for items. Participants needed to identify how many spatial subgroups/clusters a specific type has. We varied the number of spatial subgroups of a target type from two to five randomly (Fig.~\ref{fig:tasks} (a)).

\vspace{2pt}
\noindent
\textbf{Compare.} \, 
\textit{What is the highest rating of the items in a specific type associated with green labels, and how many items exist at the rating?}
One target type existed with green labels, while the other labels were gray. All items were colored white. Participants compared the ratings of subgroups with the target type and counted items in the highest-rated subgroup. We randomly varied the number of spatial subgroups with the target type from 2 to 5. \hl{All items within a subgroup share the same rating}, assigned randomly between two and five (Fig.~\ref{fig:tasks} (b)).
% One target type existed with green labels, while other labels were gray. All items were colored white. Participants compared the ratings of subgroups with the target type and counted items in the highest-rated subgroup. We randomly varied the number of spatial subgroups with the target type from 2 to 5. Each subgroup had the same rating, assigned randomly between two and five (Fig.~\ref{fig:tasks} (b)).

\vspace{2pt}
\noindent
\textbf{Summarize.} \, 
\textit{In the two-colored groups, which group has a higher average rating?}
We selected two target types: items and labels in one target were red, and those in the other were blue. The rest were gray. Participants calculated the average rating for each target type and compared them. Each target type had two to three randomly assigned subgroups, where \hl{all items within a subgroup share the same rating}, assigned randomly between two and five (Fig~\ref{fig:tasks} (c)).

The order of tasks was fixed by increasing task difficulty: Identify, Compare, and Summarize. We divided the participants into 3 groups and counterbalanced the order of label placement methods using a Latin square in each task.

\subsection{Participants \& Procedures}
We recruited 15 participants (male = 9, female = 6, age = 20-31, VR/AR experience = 6, no colorblindness) through a university mailing list. Each participant completed 45 trials: 3 label placement conditions $\times$ 5 trials $\times$ 3 tasks. Participants took an average of 11.2 minutes to complete the study, and they were offered a \$10 gift card as compensation. 

Participants were asked demographic questions and completed a consent form. We then introduced the study and tasks and asked participants to complete the tasks as accurately and as fast as possible. 
Participants had training sessions for label placement methods and tasks to become familiar with them.
Participants could take a break between each trial. After finishing the study, we collected participants' mental load for label placement methods by using a standard NASA TLX questionnaire. 
% In addition, they completed a qualitative questionnaire, including five questions for each label placement method. The five questions were: Q1. Was the layout easy to use? Q2. Was the layout easy to learn? Q3. Will you be able to use the layout in the future? Q4. Was the layout reasonable? Q5. Was the layout confusing? 
Lastly, we asked them to rank three label placement methods based on their preference.

% here is a procedure: filling out demographics info and a consent form, introducing a study. Before each task, the instructor will introduce the task and reminded participants to perform the task as precisely and as quickly as possible during the timed trials. Participants will be encouraged to spend as much time as needed on training. They can take breaks between each trial and task. After each task, the instructor will collect participants’ comments. After the final task, participants will fill out a qualitative questionnaire.

% \subsection{Data Analysis Methodology}

% Each participant will complete 45 study trials: 3 label conditions $\times$  5 trials $\times$ 3 tasks.

% For each trial, we randomly assign names and ratings to groups of items. We create a virtual scene that fits into the target AR device's field of view. We divide the scene into 5 zones and place items randomly into each zone. Each zone has the same number of items. Additionally, each zone doesn't contain more than two spatial clusters with the same item name.
\section{Results}

\vspace{5pt}
\noindent  
\textbf{Accuracy and completion time.} \,
Participants completed three tasks with similar accuracy (Situated Individual (SI): $91.60\% \pm 3.66\%$ (95\% confidence interval), Situated Grouped (SG): $90.70\% \pm 3.83\%$, and Centered Grouped (CG): $92.00\% \pm 3.57\%$). We analyzed the accuracy of participants by using a generalized linear mixed model, and there was no statistically significant effect of label placement methods on accuracy. 

\begin{figure}[t]
  \centering
  \includegraphics[width=0.75\linewidth]{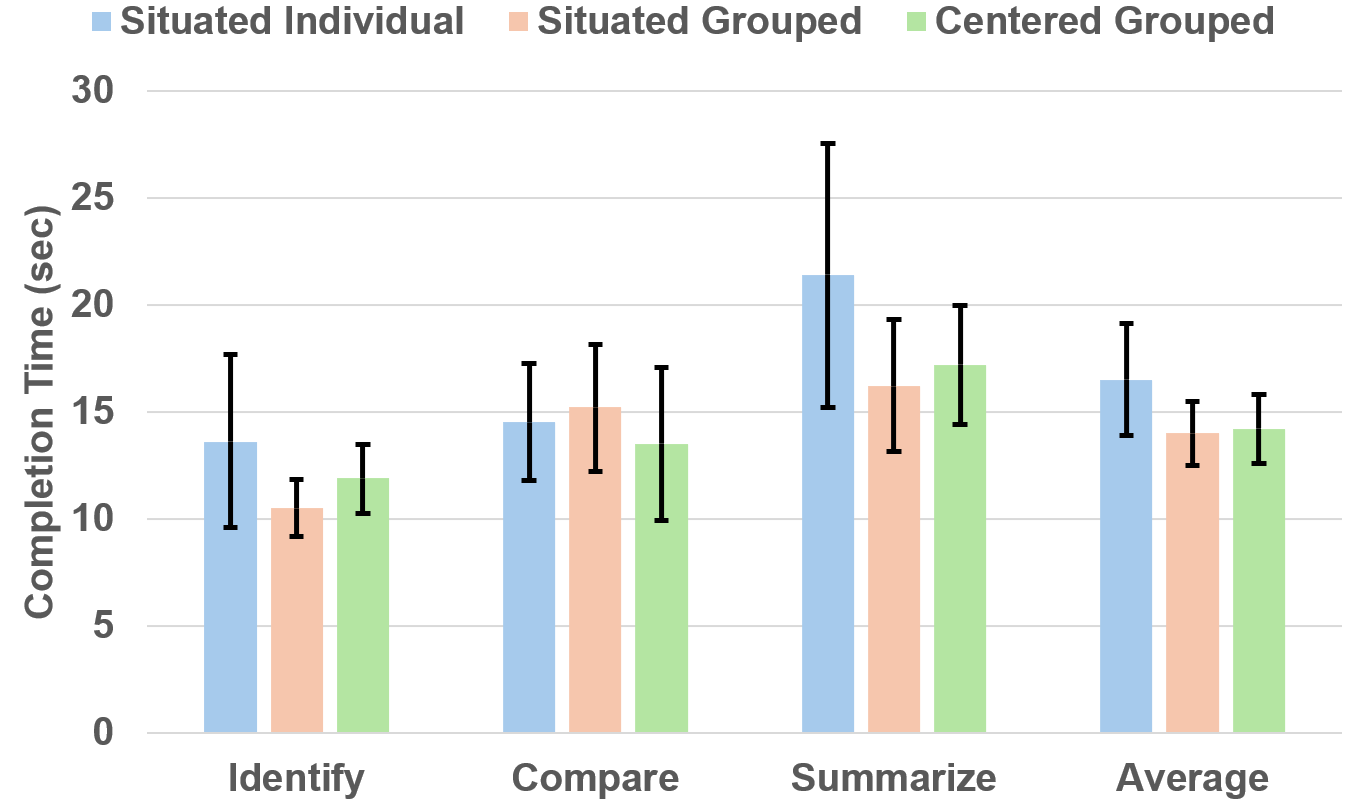}
   \vspace{-4 mm}
  \caption{%
  Mean completion time with 95\% confidence intervals for each task and all the tasks.
  }
  \vspace{-4 mm}
  \label{fig:completiontime}
\end{figure}

Fig.~\ref{fig:completiontime} shows the completion time of participants in different tasks using three label placement methods with $95\%$ confidence intervals.
We analyzed participants' completion time by using ANOVA. We found that there were significant differences among label placement methods ($p<0.05$). The Tukey p-value adjustment method was used for post-hoc analysis. Participants performed the tasks significantly faster ($p<0.05$) with SG ($14.0s \pm 1.51s$) than with SI ($16.5s \pm 2.61s$). There was no statistically significant difference between SG and CG ($14.2 \pm 1.61s$) on completion time. Additionally, there was no significant interaction between label placement methods and tasks on accuracy and completion time.

\begin{figure}[t]
  \centering
  \includegraphics[width=\linewidth]{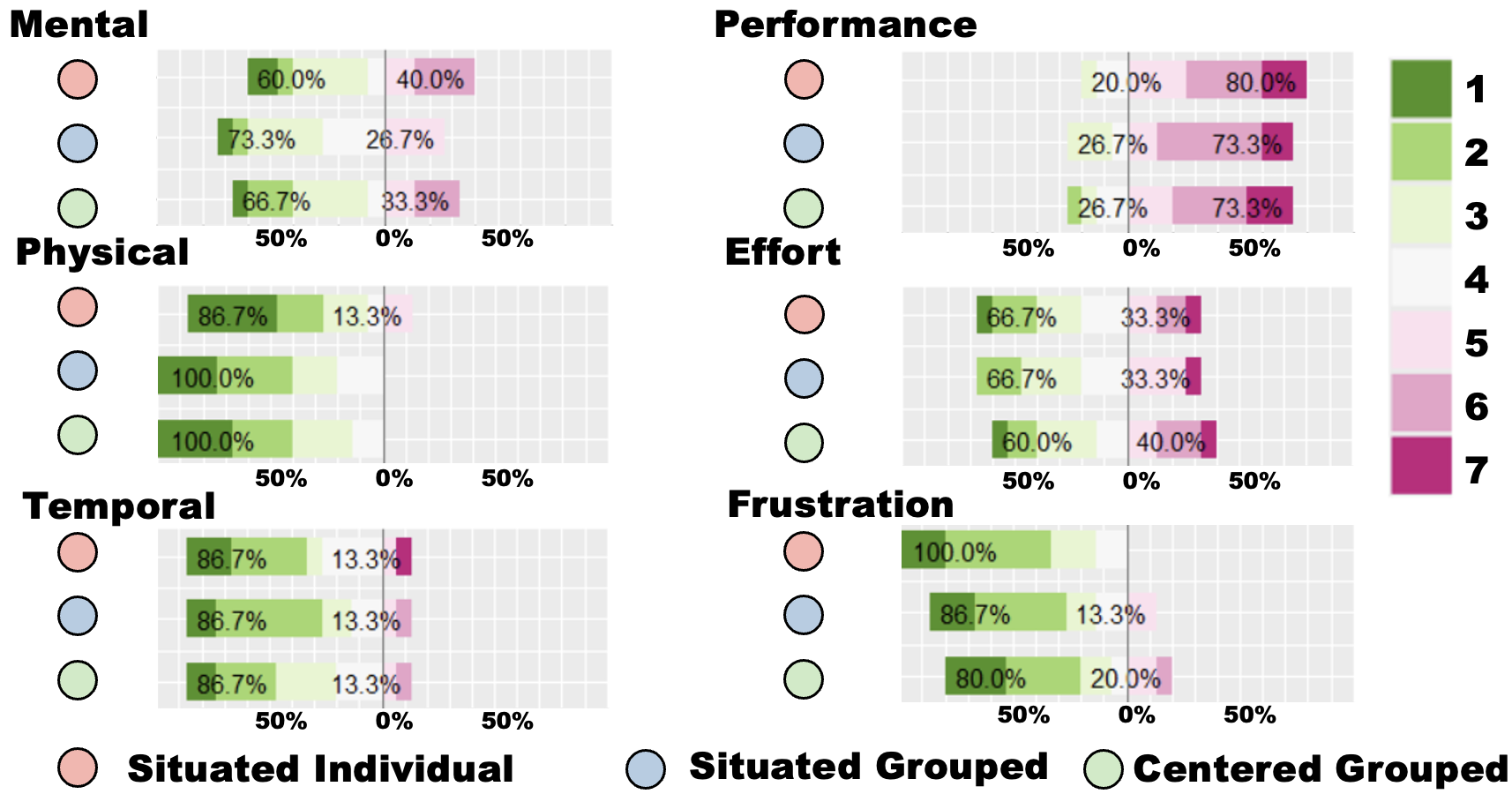}
  \vspace{-7 mm}
  \caption{%
    NASA-TLX workload ratings of comparing three layouts. The percentage of low workload (green) and high workload
ratings (pink) are shown within the bars.
  }
   \vspace{-4 mm}
  \label{fig:nasatlx}
\end{figure}

\vspace{5pt}
\noindent
\textbf{NASA-TLX workload.} \,
% (Intercept)     4957268   1 1181.0000 <2e-16 ***
% Layout              131   2    0.0156 0.9846    
% Workload         528679   5   25.1901 <2e-16 ***
% Layout:Workload    6136  10    0.1462 0.9990    
Fig.~\ref{fig:nasatlx} describes the results of participants' ratings on their mental workload for three layout placement methods.
The mean ratings are SI ($3.43 \pm 0.38$), SG ($3.38 \pm 0.34$), and CG ($3.45 \pm 0.37$). To analyze the ratings, we used a Friedman test, and the results showed that there was no statistically significant difference among the placement methods.

% Figure 14 shows the participants’ subjective ratings on
% the mental load of using different methods in different tasks after each
% trial, where 1 means low and 7 means high mental load. The respective
% mean ratings are NO = 2.55 (�� = 1.1), FORCE = 2.74 (�� = 1.29), and
% OURS = 2.36 (�� = 1.08) on NBA, and NO = 3.83 (�� = 1.59), FORCE =
% 3.02 (�� = 1.27), and OURS = 3.01 (�� = 1.20) on STU. Since a Shapiro-
% Wilk test shown that the ratings did not follow a normal distribution, we
% used a Friedman test with a null hypothesis that the participants had the
% same mental load with the three methods. Findings revealed significant
% differences in the Compare (�� = .0071) task on NBA and in both the
% Identify (�� = .0004) and Compare (�� < .0001) tasks on STU.
\vspace{5pt}
\noindent
\textbf{Participants' preferences and feedback.} \,
8, 4, and 3 out of 15 participants preferred the CG, SI, and SG methods for label placement, respectively. 
% \hl{Meanwhile, 5, 4, and 5 participants considered these methods their least preferred.} 
There was no significant difference when testing the preference for label placement methods by using a chi-squared test.
Participants who preferred the CG method expressed that it is clear, simple, and less crowded and that labels are close to items (\hl{P11: ``clearly see where the groups were, and there were not too many labels in the way''}). 
\hl{However, 3 participants expressed that the CG method is confusing because labels are too close to items.}

Some participants like the SI method because there is a one-to-one mapping between items and labels (\hl{P10: ``easily/quickest to find items''}) \hl{while 7 participants expressed that there are many labels}. 
Lastly, participants mentioned that the SG method is clear to view labels because labels are grouped, and there are no overlaps between labels and items (\hl{P2: ``The layout is very clear''}) \hl{while one participants said ``Having the label far away from the objects made it a little bit harder to count the objects}''.
% %[LET'S COMBINE IT WITH ACCURACY& COMPLETITION TIME, AND REFERENCE, AND REMOVE FEEDBACK]

%completion time
% Error: Within
%              Df  Sum Sq  Mean Sq F value Pr(>F)  
% TASK          2 0.00182 0.000910   0.451 0.6378  
% LAYOUT        2 0.01309 0.006544   3.247 0.0424 *
% TASK:LAYOUT   4 0.00275 0.000688   0.342 0.8494  
% Residuals   117 0.23583 0.002016   

 % contrast          estimate     SE  df t.ratio p.value
 % LAYOUT0 - LAYOUT1  -0.1695 0.0667 117  -2.541  0.0329
 % LAYOUT0 - LAYOUT2  -0.0962 0.0667 117  -1.441  0.3231
 % LAYOUT1 - LAYOUT2   0.0733 0.0667 117   1.099  0.5166

% 1 0       16.5   225
% 2 1       14.0   225
% 3 2       14.2   225

% L0: 13.89471 19.12365
% L1: 12.47532 15.49041
% L2: 12.58563 15.79870

%preference:
%X-squared = 2.8, df = 2, p-value = 0.2466

% \subsection{Accuracy}

% \subsection{Completion Time}

% \subsection{Mental Load}
% TODO:!!!

% \subsection{Comments}

\section{Discussion}

\vspace{5pt}
\noindent
\textbf{Use the Grouped methods.} \,
Using a single label for spatially grouped items helped participants complete tasks faster than using a label for each item, supporting the H1 hypothesis.   
Although participants mentioned that the Situated Individual is easy to use and learn, it causes visual clutter without providing extra information. One participant said, ``The Situated Individual layout method provides too much information to attempt to process quickly (P9)''. 
It was found that visual clutter increases response times during visual search tasks~\cite{Beck:2010}.
In contrast, the Grouped methods assisted participants in obtaining information related to target labels. 
Additionally, participants preferred grouping methods (Situated Grouped: 3 participants, Centered Grouped: 8 participants) because they could easily find groups associated with target labels (P11).

\vspace{5pt}
\noindent
\textbf{Balance between the lengths of leader lines and visual clutter.} \,
Participants performed the tasks similarly when they used the Situated Grouped method and the Centered Grouped method, rejecting the H2 hypothesis.
Even though the distance between a representative label and its associated items is shorter in the Centered Grouped method, sometimes a label partially blocks an item(s). Moreover, in some cases, proximity between labels and objects in the Centered Grouped method causes visual clutter in subgroups, so users may have difficulty finding information. This may cause a similar completion time for both methods. Therefore, unlike environments with sparsely distributed items where short leader lines are preferred~\cite{Vollick:2007}, in our target scenes, we need to balance between the leader line lengths and visual clutter between labels and objects.  

\vspace{5pt}
\noindent
\textbf{Limitations.} \,
In this study, we focused on items within users' FOV. However, in real applications, items of interest can be located outside users' FOV~\cite{Lin:2023}. Future research should investigate how labels of groups of out-of-view items are displayed.
Although we tried to minimize overlap between items and labels by using a force-directed method in the Centered Grouped method, sometimes labels partially occlude items. A possible solution is to make labels transparent because there is little space to move labels in visually cluttered scenarios.
We will explore the effect of transparency and colors of the labels with different real-world scenes.
Another limitation of the study is that simulating AR in VR may not completely represent real environments, although we followed existing experiments~\cite{Lin:2023,Chen:2023}. Furthermore, we used simple backgrounds and objects of the same size and shape to focus on label placement methods, following the previous work~\cite{Lin:2023}.
Additionally, while our study investigated label placement methods for a total of 50 items within users' FOW, a scenario with more items may arise.
\hl{
Currently, users need to manually adjust the distance thresholds for spatial grouping. We will explore automatically calculating these thresholds based on the level of visual clutter and the size and shape of objects, to scale to real AR settings with irregular object shapes or more dynamic environments.} 
Lastly, we focused on two locations of representative labels for the grouped items: above and center. However, a representative label can be located in other places, such as the left or right of a group. We will explore other possible locations for the labels. 
% TODO: embedded labels??\\

%%TODO:Using cubes and a simple background 

% situations one - vertical move.
% object placement - no overlapping - maybe cause an issue for grouped one.
% (centered grouped - force-directed method: minimizing overlapping: sometimes occludes objects, how can we solve this issue? transparent or color? without hurting information recognition??)

\section{Conclusion}
\hl{We have presented a study on how label placement affects a visually cluttered environment in AR. We compared three label placement methods in a scene where multiple items of the same type are spatially close in the user's field of view. Our findings indicate that using a label for spatially grouped items helps users to identify, compare, and summarize data in terms of completion time. 
% Additionally, many participants preferred a single label for spatially grouped items of the same type.

}

 % We investigate methods for placing labels in AR environments that have visually cluttered scenes.
 % As the number of items increases in a scene within the user' FOV, it is challenging to effectively place labels based on existing label placement guidelines. To address this issue, we implemented three label placement techniques for in-view objects on the AR screen. We specifically target a scenario, where various items of different types are scattered within the user's field of view, and multiple items of the same type can be situated close together.      
 % We evaluate three placement techniques for three target tasks. Our study shows that using a label for spatially grouped the same types of items is beneficial for identifying, comparing, and summarizing data. 

% The authors wish to thank A, B, and C. This work was supported in part by
% a grant from XYZ.}

%\bibliographystyle{abbrv}
\bibliographystyle{abbrv-doi}

\bibliography{template}
\end{document}